\documentclass[apl,preprint,superscriptaddress]{revtex4}

\usepackage{amsmath, amsfonts, amssymb}
\usepackage{graphicx}

\begin{document}

\title{Measurement of light diffusion in ZnO nanowire forests}

\author{Marijn A. M. Versteegh}
\affiliation{Debye Institute for Nanomaterials Science, Utrecht University, Princetonplein 1, 3584 CC Utrecht, The Netherlands}
\author{Ruben E. C. van der Wel}
\affiliation{Debye Institute for Nanomaterials Science, Utrecht University, Princetonplein 1, 3584 CC Utrecht, The Netherlands}
\author{Jaap I. Dijkhuis}
\affiliation{Debye Institute for Nanomaterials Science, Utrecht University, Princetonplein 1, 3584 CC Utrecht, The Netherlands}

\begin{abstract}
Optimum design of efficient nanowire solar cells requires better understanding of light diffusion in a nanowire array. Here we demonstrate that our recently developed ultrafast
all-optical shutter can be used to directly measure the dwell time of light in a nanowire array. Our measurements on disordered ZnO nanowire arrays, ``nanowire forests,'' indicate that
the photon mean free path and the dwell time of light can be well predicted from SEM images.
\end{abstract}

\maketitle Nanowire solar cells are promising devices for solar energy conversion \cite{garnett 2011}. They are inexpensive because of the small amount of material needed, and efficient because
of strong light absorption and rapid carrier collection. In the past few years, photovoltaic devices have been reported based on arrays of dye-sensitized ZnO nanowires
\cite{baxter 2005,law 2005}, nanoflowers \cite{jiang 2007}, nanotrees \cite{ko 2011}, silicon nanowires \cite{tsakalakos 2007, garnett 2010}, nanocones \cite{lu 2010}, and
microwires \cite{kelzenberg 2010}, CdS nanopillars \cite{fan 2009}, and multi-layered nanopillars \cite{naughton 2010} and nanorods \cite{kuang 2011}.

The enhanced optical absorption of these photovoltaic devices compared to planar solar cells can be attributed to three effects. First, the reflectivity is reduced \cite{hu
2007, muskens 2008, zhu 2009, kupec 2010}. Second, nanowires capture and confine incoming light into guided modes, leading to concentration of the electromagnetic field inside
the absorbing material \cite{naughton 2010, kupec 2010, zhang 2009, lin 2009, cao 2010}. Third, light travels a long diffusive path through the nanowire array, as a result of
multiple scattering between the wires. A photon not converted to an electron-hole pair in a first nanowire hit, gets many additional chances.

This diffusion effect has been experimentally studied by static optical reflection and absorption measurements and current-voltage characterization of solar cells
\cite{garnett 2010, muskens 2008}. For a good understanding of the light diffusion dynamics, however, also time-resolved measurements are needed. Powerful interferometric
methods have been developed to measure the transport of light through strongly scattering media \cite{kop 1997, johnson imhof 2003}. Application of these methods is, however,
quite difficult, because of the advanced optical setup needed, the speckle by speckle measurement, and the complicated data analysis.

In this Letter, we present a simple method to directly measure the diffusive dwell time of light in a disordered ZnO nanowire array, a ``nanowire forest.'' The measured dwell
times of light turn out to be reasonably well accounted for by a simple ray-optics diffusion model. In this model, the photon mean free path (MFP) is estimated from scanning
electron microscope (SEM) images of the nanowire forests. Consequently, SEM images can be used to predict the light diffusion in this type of samples, and to optimize the
solar cell performance.

For the measurements of the dwell time, we use our recently developed ultrafast all-optical shutter \cite{versteegh shutter}, with a ZnO nanowire forest replacing the ZnO
single crystal. Two 125-fs laser pulses, a strong 800-nm gating pulse and a weaker probe pulse with a wavelength between 385 and 415 nm, are sent through the ZnO nanowire
forest with a variable delay. If the delay is such that the pulses are simultaneously present inside the sample, two-photon absorption of a probe photon and a gating photon
occurs. Measurement of the probe transmission as a function of delay reveals the time photons spend inside the sample.

\begin{figure}
\begin{center}
\includegraphics[width=0.7\textwidth]{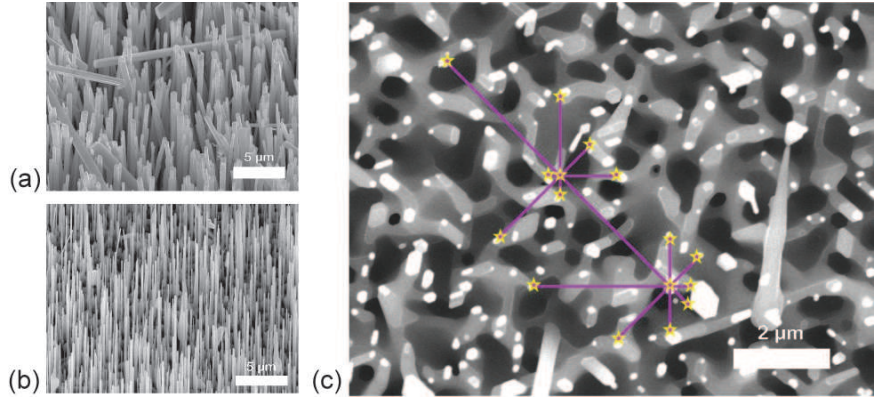}
\caption{(color online) SEM images of the nanowire forests. (a) Nanowire forest 1. (b) Nanowire forest 2. (c) Top view of nanowire forest 2. Rays and scattering events are shown, indicating how we determined the photon mean free path from this image.}
\end{center}
\end{figure}

We have performed measurements on two disordered ZnO nanowire arrays (Fig. 1). Note that disorder in the position of nanowires has been shown to enhance the absorption of
light \cite{bao 2010}. The nanowire forests were epitaxially grown on sapphire substrates using the carbothermal reduction method with gold particles as catalysts
\cite{prasanth 2006}. In the furnace, first a porous ZnO seed film grows on the sapphire crystal. On top of that film the nanowires emerge, with their \textit{c}-axes parallel
to the wire. The nanowire forests therefore consist of a nanowire part and a film part. The film is well visible in the top view image of forest 2 (Fig. 1c). The selected
nanowire forests differ from each other with respect to nanowire length, diameter, and density. Properties of the samples are summarized in Table 1.

\begin{table*}
\begin{center}
Table 1. Properties of the samples examined in this work,
determined from SEM images. \vspace{3mm}\\
\centerline{\begin{tabular}{c c c c c c c c c}
\hline
\hline
\noalign{\vspace{0.5mm}}
\small
Sample &Nanowire &Average &Nanowire &ZnO fraction &Photon MFP  &Film   &ZnO fraction &Photon MFP         \\
       &length  &diameter &density  &nanowire part  &nanowire part     &thickness  &film part     & film part     \\
\hline
\noalign{\vspace{0.5mm}}
Nanowire forest 1  & 20 $\mu$m & 250 nm & 0.85 $\mu$m$^{-2}$ & 0.080  & 5.9 $\mu$m & 1.5 $\mu$m & 0.75 & 0.54 $\mu$m  \\
Nanowire forest 2 & 7.6 $\mu$m & 170 nm & 2.4 $\mu$m$^{-2}$ & 0.086 & 2.1 $\mu$m & 1.5 $\mu$m & 0.75 & 0.54 $\mu$m \\
\hline
\hline
\end{tabular}}
\end{center}
\end{table*}

\begin{figure}
\begin{center}
\includegraphics[width=0.4\textwidth]{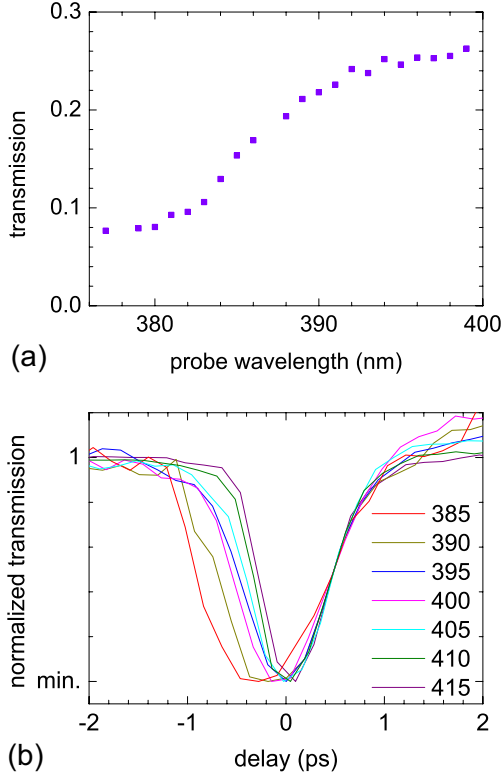}
\caption{(color online) Experimental results on nanowire forest 1. (a) Absolute transmission. (b) Transmission of the probe pulse versus delay with respect to the gating pulse. Positive delay means that the probe pulse arrives after the gating pulse. Probe wavelengths are indicated in nm. The shape and width of the dip turned out to be independent of gating fluence. For clarity, the initial transmission level and the transmission minima are normalized. For shorter probe wavelengths the two-photon absorption dip is wider, revealing a longer dwell time of the probe photons inside the sample.}
\end{center}
\end{figure}

Experimental results on nanowire forest 1 are shown in Fig. 2. Figure 2a shows the measured transmission of the probe pulse versus wavelength in the absence of a gating pulse.
The transmitted light was diffusive; no beam was present behind the sample. The measured probe transmission in the presence of a gating pulse is presented
in Fig. 2b. Clearly, when the two pulses arrive approximately at the same time, the transmission of the probe pulse is reduced: the effect of two-color two-photon absorption.
We also observe that the two-photon absorption dip is wider for shorter probe wavelengths, like in Ref. \cite{versteegh shutter}, which is explained by the smaller group
velocity at wavelengths close to the exciton resonance in ZnO.

The sample of Ref. \cite{versteegh shutter} is a 523 $\mu$m thick single crystal, through which light propagates in a straight line, while here we have a 21.5 $\mu$m
thick nanowire forest, through which light propagates in a diffusive manner. If the photons would have traversed the forest in a straight line, then the two-photon absorption dip
should be about 1/24 of the widths measured in Ref. \cite{versteegh shutter}. The dips in Fig. 2b are, however, much wider, around 1 ps. This directly shows that diffusion severely
lengthens the dwell time of the photons, which is in agreement with our direct observation that the transmitted light is diffusive. The dwell time of photons inside the
nanowire forest apparently is in the order of 1 ps.

For further analysis of our experimental results, we used the following simple ray-optics diffusion model. Since the incoming gating and probe photons propagate parallel to
the nanowires, initially scattering and diffusion are not very strong. There is waveguiding through the wires and propagation through the air between the wires. In our rough
model, all photons travel ballistically to the center of the forest, where they scatter, and diffusion commences. The diffusion is modeled using standard diffusion theory
\cite{jost 1960} with an isotropic MFP for the nanowire part and a different isotropic MFP for the film part of the sample. We have set the boundary conditions such that a
photon escapes as soon as it reaches the top surface of the forest or the interface between the porous film and the sapphire.

The photon MFP is estimated from the SEM images. It is assumed that whenever a photon hits a nanowire, it scatters with equal probability in all directions. From many
nanowires on our SEM images we have drawn rays in eight directions, as indicated for one wire in Fig. 1c. The photon MFP in the nanowire part equals the average distance to
the next nanowire determined in this way, multiplied by $\sqrt{2}$ to account for motion in the vertical direction. For the film part the MFP is estimated in a similar manner.
The results are given in Table 1. In this simple model, the MFP is taken to be independent of wavelength.

From the photon MFP the diffusion coefficient is calculated using $D=cl/[3n_{g\textrm{ eff}}(\lambda)]$ \cite{jost 1960}, where $c$ is the vacuum speed of light, $l$ is the
photon MFP, and $n_{g\textrm{ eff}}$ is the wavelength-dependent effective group index of refraction, given by $ n_{g\textrm{
eff}}(\lambda)=n_{\textrm{eff}}(\lambda)-\lambda\mathrm{d}n_{\textrm{eff}}(\lambda)/\mathrm{d}\lambda$. Here, the effective index of refraction $n_{\textrm{eff}}$ is
determined from
$n_{\textrm{eff}}=\mathrm{Re}[\sqrt{\varepsilon_{\textrm{eff}}(\lambda)}]=\mathrm{Re}[\sqrt{f\varepsilon_{\textrm{ZnO}}(\lambda)+(1-f)\varepsilon_{\textrm{air}}}]
$\cite{kirchner 1998, diedenhofen 2010}, where $\varepsilon_{\textrm{ZnO}}(\lambda)$ is the complex dielectric constant of ZnO, $\varepsilon_{\textrm{air}}$ is the dielectric
constant of air, and $f$ is the ZnO fraction, given in Table 1. The real part of the dielectric constant of ZnO is taken from Refs. \cite{jellison 1998, yoshikawa 1997, bond
1965}, the imaginary part from Ref. \cite{versteegh prb}. Apart from diffusion, absorption also influences the dwell time of light inside the sample. We included in our model
a linear absorption factor derived from the experimental data shown in Fig. 2a. The effect of three-photon absorption of the 800-nm gating pulse on the dwell time is
neglected, as we do not observe any change in shape of the dip for increasing gating fluence, and thus for increasing three-photon absorption.

\begin{figure}
\begin{center}
\includegraphics[width=0.4\textwidth]{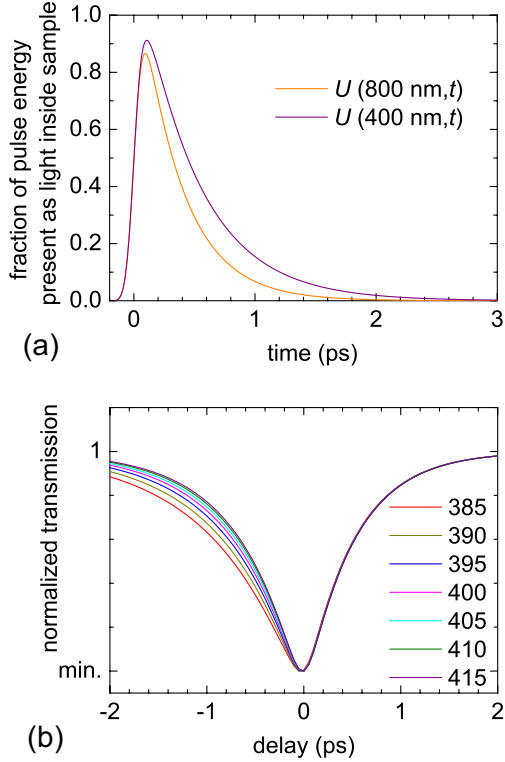}
\caption{(color online) Results from our ray-optics diffusion model, with the photon MFP derived from the SEM images. (a) Dynamical fraction of the pulse energy present as light inside forest 1, calculated for an 800-nm pulse and for a 400-nm pulse. The increase is due to the incoming pulse. The decrease is due to two effects: photons leaving the sample and absorption. (b) Calculated transmission through forest 1 versus delay for various probe wavelengths, indicated in nm.}
\end{center}
\end{figure}

Using our diffusion model, we calculated the functions $U(\lambda,t)$, defined as the wavelength- and time-dependent fraction of the pulse energy present as light inside the
sample. The transmission versus delay $\delta$ of the probe with respect to the gate pulse can subsequently be calculated from
\begin{equation}
T(\lambda_{\textrm{probe}},\delta)=1-c_1\int_{-\infty}^{\infty}U(\lambda_{\textrm{probe}},t-\delta)U(\lambda_{\textrm{gate}},t)\mathrm{d}t,
\end{equation}
where $c_1$ is a constant. The integral describes the two-photon absorption of a probe photon and a gate photon. Results for forest 1 are presented in Fig. 3.

In Fig. 4 the two-photon dip widths of Fig. 3b are compared with the experimental dip widths of Fig. 2b. On average, the dip width is well described by our simple diffusion
model. Also for nanowire forest 2, the model dip widths reasonably agree with the measured ones. The dashed lines in Fig. 4 indicate the average photon dwell time in the two
samples, which is simply given by
\begin{equation}
\bar{\tau}(\lambda)=\int_{-\infty}^{\infty}U(\lambda,t)\mathrm{d}t.
\end{equation}
As is clear from Fig. 4, the average dwell time is close to the dip width.

\begin{figure}
\begin{center}
\includegraphics[width=0.4\textwidth]{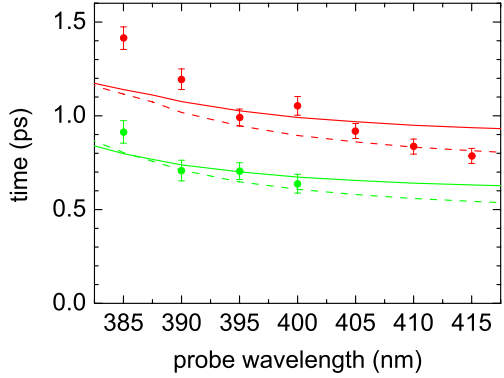}
\caption{(color online) Comparison between diffusion model and optical shutter measurements. Red points: measured two-photon absorption dip width (full-width half-maximum) of forest 1 (Fig. 2b). Red solid line: dip width (full-width half-maximum) of forest 1 according to the diffusion model (Fig. 3b). Red dashed line: average time spent inside forest 1 by a probe photon as calculated from the diffusion model. Green points and lines: idem for forest 2.}
\end{center}
\end{figure}

The model results show larger dip widths for shorter wavelengths, in agreement with the experimental data. This is the result of the higher group index of refraction and thus
the smaller diffusion coefficient. The slopes in Fig. 4, however, deviate from the experimental results, due to our rough approximation that the MFP is independent of
wavelength. A more advanced model could take the wavelength-dependency into account. We wish to emphasize however, that it is of practical use that a simple ray-optics model,
in which the photon MFP is measured from SEM images, describes the light diffusion and the dwell times of light already reasonably well.

In conclusion, we have demonstrated a simple and straightforward method to measure the diffusive dwell time of light inside ZnO nanowire forests. This method is based on
two-photon absorption of a gating photon and a probe photon. The diffusion can be reasonably well described by standard diffusion theory, in combination with a simple
ray-optics model, where the photon MFP is determined from SEM images. We expect these results to be of value for nanowire solar cell research.

We thank D. H. van Dorp, H. Y. Li, and D. A. M. Vanmaekelbergh for the nanowire forests, C. R. de Kok and P. Jurrius for technical support, and R. E. I. Schropp, M. di Vece,
and Y. Kuang for comments.

\end{document}